\documentclass{PoS}

\title{The camera system for the IceCube Upgrade}

\ShortTitle{The camera system for the IceCube Upgrade}

\author{
The IceCube Collaboration\footnote{For collaboration list, see PoS(ICRC2019)1177}\\
{\itshape \href{http://icecube.wisc.edu/collaboration/authors/icrc19_icecube}{http://icecube.wisc.edu/collaboration/authors/icrc19\_icecube}}\\
E-mail: \email{wkang@icecube.wisc.edu, christoph.toennis@gmx.de, rott@skku.edu}
}

\abstract{
The IceCube Neutrino Observatory is a cubic kilometer volume neutrino detector installed in the Antarctic at the geographic South Pole. Neutrinos are detected through the observation of Cherenkov light from charged relativistic particles generated in neutrino interactions, using an array of 86~strings of optical sensor modules. Currently an upgrade to the IceCube detector is in preparation. This IceCube Upgrade will add seven additional strings with new optical sensors and calibration devices. A new camera system is designed for this upgrade to be installed with the new optical modules. This camera system will study bulk ice properties and the refrozen ice in the drill hole. The system can also be utilized to provide information on the detector geometry including location and orientation of the optical modules and cables that can be used to calibrate IceCube Monte Carlo simulations. A better understanding of the refrozen ice in the drill hole including the complementary knowledge of the optical properties of the surrounding glacial ice will be obtained by surveying and analyzing the images from this system. The camera system consists of two types of components: an image sensor module and an illumination module. The image sensor module uses a CMOS image sensor to take pictures for the purpose of calibration studies. The illumination module emits static, monochromatic light into a given direction with a specific beam width and brightness during the image taking process. To evaluate the system design and demonstrate its functionality, a simulation study based on lab measurements is performed in parallel with the hardware development. This study allows for the development of the preliminary image analysis tool for the system. We present the prototype of the camera system and the results of the first system demonstrations.\\

\vspace{4mm}
{\bfseries Corresponding authors:}
Woosik Kang$^{1}$, \speaker{Christoph T\"onnis}$^{1}$, Carsten Rott$^{1}$ \\
{$^{1}$ \itshape Department of Physics, Sungkyunkwan University, Suwon 16419, Korea}\\
}

\FullConference{36th International Cosmic Ray Conference -ICRC2019-\\
		July 24th - August 1st, 2019\\
		Madison, WI, U.S.A.}

\begin{document}

\section{Introduction}\label{sec:introduction}

\indent The IceCube Neutrino Observatory~\cite{ICdetector}, located at the geographic South Pole, is the worlds largest neutrino telescope. The observatory consists of a Cherenkov radiation detector of one cubic kilometre volume using the ultra-pure Antarctic ice~\cite{SPICE} at depths between 1.45~km and 2.45~km, as well as a square kilometer air-shower detector at the surface of the ice~\cite{IceTop}. The primary scientific objectives of the detector are the measurement of high-energy astrophysical neutrino fluxes and determining the sources of these fluxes~\cite{HEnu_Science,TXS}. Further goals include the measurement of neutrino oscillation parameters~\cite{NuOsc}, cosmic ray studies~\cite{cosmic_ray}, and search for dark matter~\cite{HESE_DM} and exotic particles~\cite{monopole}.\\ 
\indent During polar season 2022/2023 an upgrade of the IceCube detector with new digital optical modules (DOMs) is scheduled to be deployed~\cite{ICRC2019:ICU}. This upgrade comprises of seven densely instrumented strings to be installed in the centre of the IceCube detector volume. On each string DOMs will be regularly spaced with a vertical separation of 3~m between depths of 2160~m and 2430~m below the surface of the ice. Three different types of DOMs will be used: The pDOM which is based on the design of the existing IceCube DOMs with upgraded electronics, the D-Egg which has two eight-inch PMTs (facing up and down respectively), and the mDOM which has 24~three-inch photomultiplier tubes (PMTs) distributed for close to uniform directional coverage. The diameters of the modules range from 35.6~cm (mDOM) to 30~cm (D-Egg). The D-Egg and the mDOM glass pressure vessels have a central straight section as needed to accommodate the PMTs. \\ 
\indent One of the main objectives of the IceCube Upgrade~\cite{ICRC2019:ICU} is to contribute to the understanding of neutrino oscillations by probing the tau sector, as part of the world-wide program to test unitarity of the PMNS matrix. Another objective is to precisely characterize the IceCube detector medium and thereby reduce the uncertainties in reconstruction of direction and energy of neutrino events throughout the full IceCube detector volume. The latter objective will be achieved using a series of improved and novel calibration devices. One key component of the IceCube Upgrade is the IceCube Upgrade camera system~\cite{ICRC2017:Gen2_camera}. This system aims to measure the refrozen ice in the drill holes for the upgrade strings and the optical properties of the ice between the strings. Additionally, information on the position and orientation of the optical sensors will be obtained. To do this the camera system utilizes three camera modules installed inside each newly installed DOM in the upgrade to measure light emitted from illumination modules that are installed next to each camera pointing in the same direction. With this setup images of both reflected and transmitted light will be taken.\\
\begin{figure}
\centering
\begin{minipage}{.8\textwidth}
\centering
\includegraphics[width=.35\linewidth]{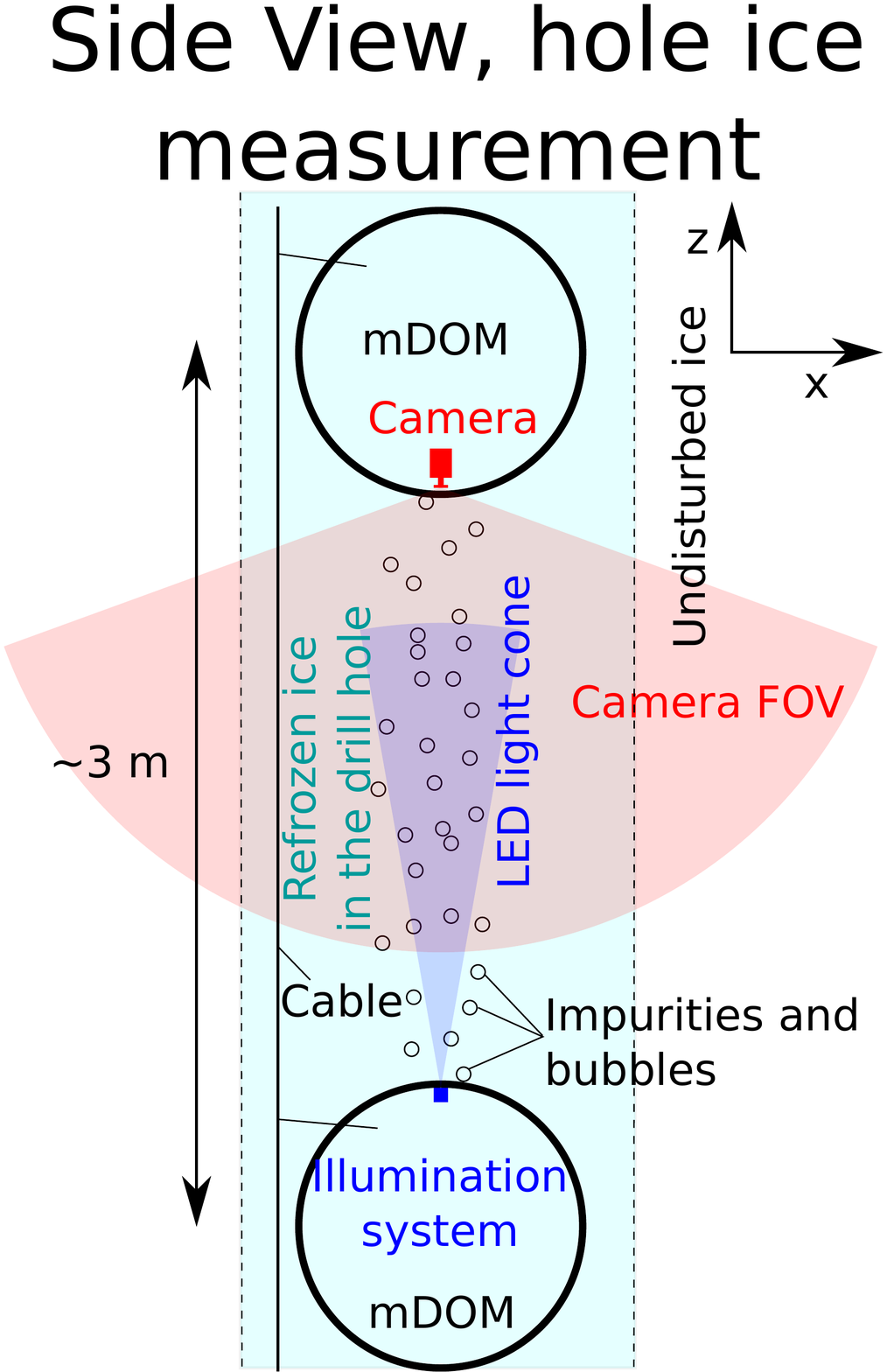}
\includegraphics[width=.47\linewidth]{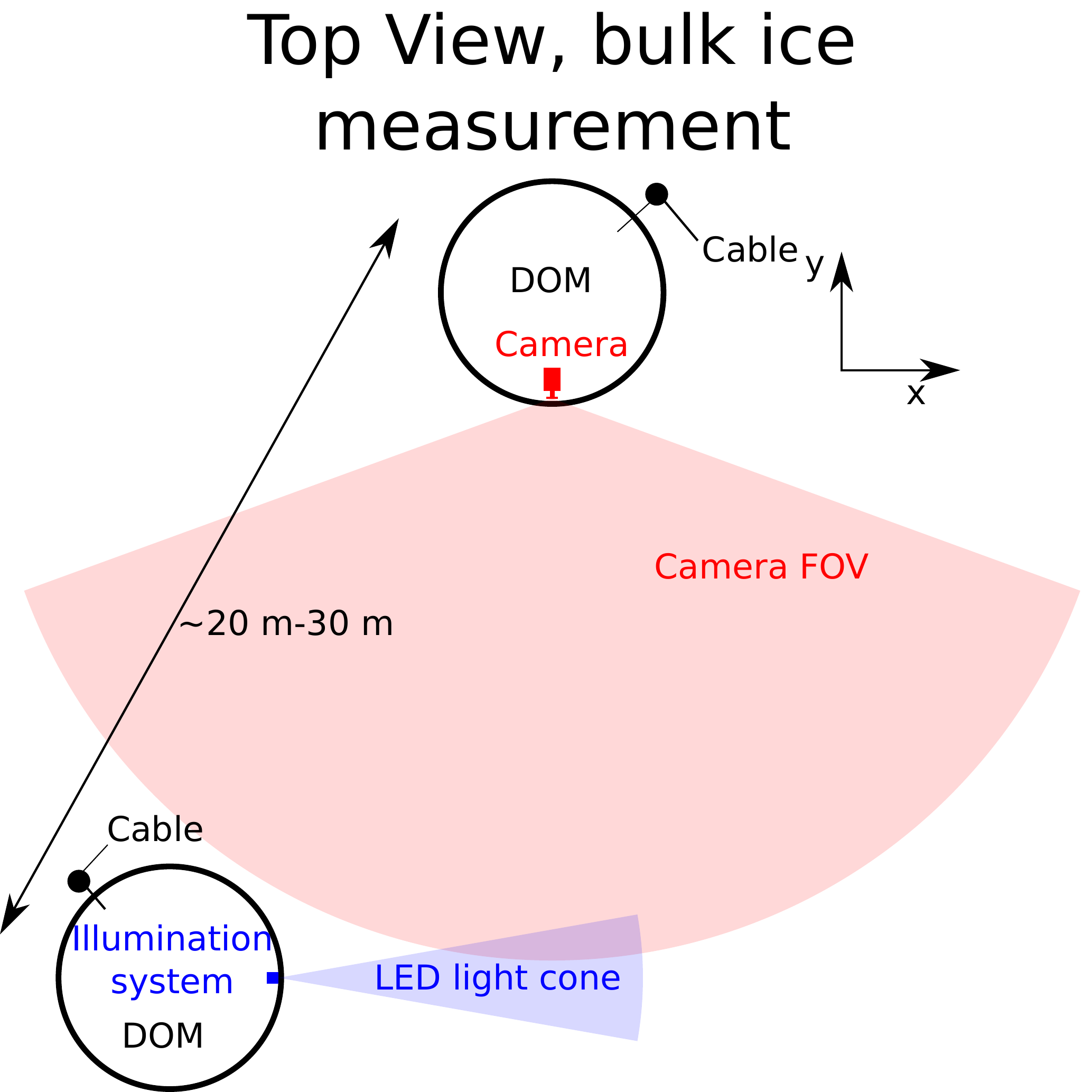}
\end{minipage}
\caption[margin=1cm]{Proposed measurements for the IceCube Upgrade camera system. Left: Refrozen hole ice measurement utilising two vertically separated optical modules on the same string. A downward facing camera observes an up-ward pointing LED from the DOM below. Right: Bulk ice measurement utilising two optical modules on separate strings. A camera is observing scattered light from an LED on an adjacent string.}
\label{fig:schematic}
\end{figure}
\indent The camera system is to be installed in all three types of DOMs for the IceCube Upgrade. In the pDOM the cameras will be installed in the upper hemisphere of the DOM above the DOM mainboard pointing upwards at 35${^\circ}$. In the D-Egg all three cameras will point in the ecliptic at 120${^\circ}$ angle to each other and will be placed above the bottom PMT. In the mDOM two cameras will be placed in the upper hemisphere at opposite sides pointing upwards at an angle of 45${^\circ}$ and the third camera will be placed at the bottom of the lower hemisphere pointing directly downwards. At the top of each mDOM one additional illumination module will be placed pointing upwards to illuminate the refrozen ice above the mDOM.\\
\indent One of the major objectives of the camera system is to measure the properties of the refrozen hole ice. As depicted in the left of Fig.~\ref{fig:schematic} the downwards facing camera captures direct and scattered light from an illumination module in the DOM below. Based on the distribution of light in the images the optical properties of the refrozen ice will be measured. From a special camera system deployed below the deepest DOM of IceCube string~80~\cite{ICdetector}, a dust deposit and a central, small region of short scattering lengths, referred to as the bubble column, was found in the refrozen hole ice~\cite{Swedish}.\\
\indent The properties of the surrounding ice can be measured by capturing scattered light from illumination modules on adjacent strings as illustrated in Fig.~\ref{fig:schematic}, right. An LED on one DOM illuminates the inter-string ice medium. In a DOM on a neighboring string, a sideways pointing camera will take images of the scattered light. The optical properties of the ice can be measured based on the distribution of incident light as shown in Sec.~\ref{sec:simulation}. Furthermore, comparing images capturing light emitted in different directions, anisotropies~\cite{anisotropy} of the optical properties of the ice can be studied. Lastly the position of the light cones in these images can be used to determine the orientation and position of the newly deployed DOMs within their respective holes.\\

\section{Design of the camera system}\label{sec:design}

\begin{figure}
\centering
\begin{minipage}{.85\textwidth}
\centering
\includegraphics[width=.45\linewidth]{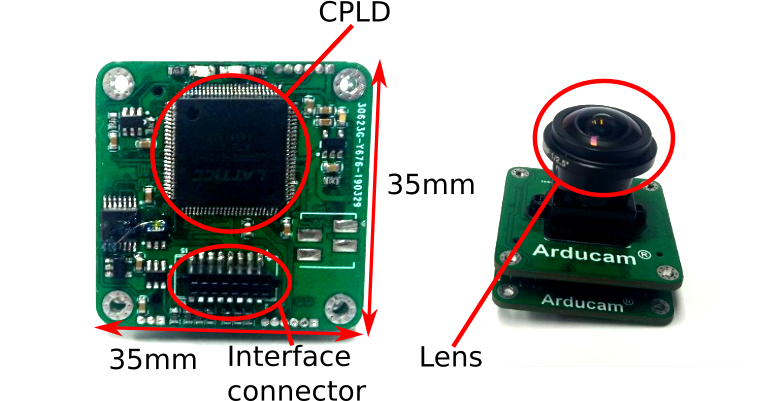}
\includegraphics[width=.45\linewidth]{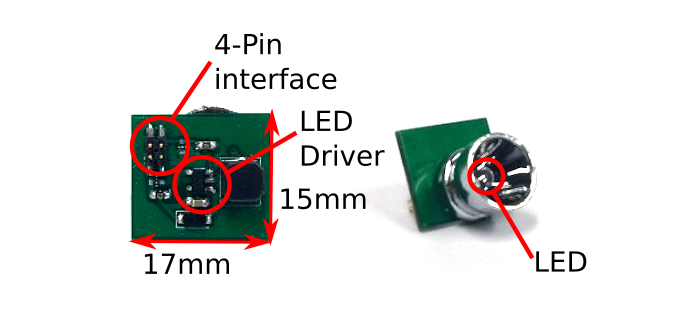}
\end{minipage}
\caption[margin=1cm]{The major components of the IceCube Upgrade camera system. Left: A photo of the camera module, consisting of a CPLD board with SPI interface connector and an image sensor board with a lens and its mount. Right: A photo of the illumination module with a LED, LED driver, and interface connector.}
\label{fig:photo}
\end{figure}
\indent The camera module uses a Complementary Metal-Oxide-Semiconductor (CMOS) image sensor\footnote{`IMX225LQR-C' manufactured by SONY Semiconductor Solutions Co.}. This image sensor is controlled by a Complex  Programmable Logic Device (CPLD)\footnote{`LCMXO2-1200HC' manufactured by Lattice Semiconductor Co.} that also handles a serial peripheral interface (SPI) communication to the mainboard and is connected to a Random Access Memory (RAM)\footnote{`MT48LC4M16A2P-6A' manufactured by Micron Technology, Inc.} on each camera board. The CPLD controls the settings of the image sensor and can set parameters such as gain or exposure time of the image sensor. Images taken by the image sensor are initially stored on the RAM, from where they can then be transferred to the surface through the CPLD and the DOM hardware and IceCube infrastructure.\\
\indent With the camera a 22~mm diameter wide-angle M12 lens with a field of view of approximately 170${^\circ}$ is used. Taking into account distortions due to the glass of the DOMs and refraction in the surrounding ice an effective field of view of approximately 120${^\circ}$ is expected, which in the mDOM will be further restricted to 90${^\circ}$ due to the holding structure in the DOM.\\
\indent The illumination module is controlled via a 4-pin interface that supplies the same 5~V operating voltage used in the camera and an enable signal that is routed through the camera and is controlled by the DOM mainboard. The light is produced by an LED\footnote{`SSL80 GB CS8PM1.13' manufactured by OSRAM Opto Semiconductors GmbH.} operating at 1~W of power consumption. The light emitted has a peak wavelength of 465~nm with a bandwidth of 25~nm and an emission cone size of 40${^\circ}$ in air. A reflective cup concentrates the light to a smaller cone, which will be 22.5${^\circ}$ in the ice. The LED is powered by an integrated LED driver circuit\footnote{`AL8860' manufactured by Diodes, Inc.}. Some technical requirements of the whole system can be seen in table~\ref{tab:ERD}.\\
\begin{table}[]
    \centering
    \begin{tabular}{|l|c|}
    \hline
        \textbf{Parameter} & \textbf{Requirement} \\
        \hline
        \hline
        Temperature & All modules have to operate at $-40^\circ C$ (Storage requirement $-50^\circ C$) \\
        \hline
        Light in detector & When inactive neither camera nor illumination module can emit any light\\
        \hline
        Current rating & The cables used in the system has to carry 0.7~A of current\\
        \hline
        Lens Field of View & The camera field of view has to be larger than $60^\circ$\\
        \hline
        Alignment & The camera alignment within the DOM has to be known within $5^\circ$\\
        \hline
        Image Sensor type & A colour image sensor has to be used for reflection photography\\
        \hline
    \end{tabular}
    \caption{A table of some of the engineering and science  requirements of the camera system.}
    \label{tab:ERD}
    \vskip-2.2ex
\end{table}
\indent Each pair of camera and illumination module is connected to the mainboard of each type of DOM via a separate cable and a 10-pin connector. Through this interface an operating voltage of 5~V is supplied and data and commands are transferred using the SPI protocol. The camera and the illumination module are enabled and disabled using the dedicated lines in the interface. The power supply and the control line for the illumination module are routed through each adjacent camera.\\ 

\section{System verification}\label{sec:verification}

\begin{figure}
\vskip-1ex
\centering
\includegraphics[width=0.8\linewidth]{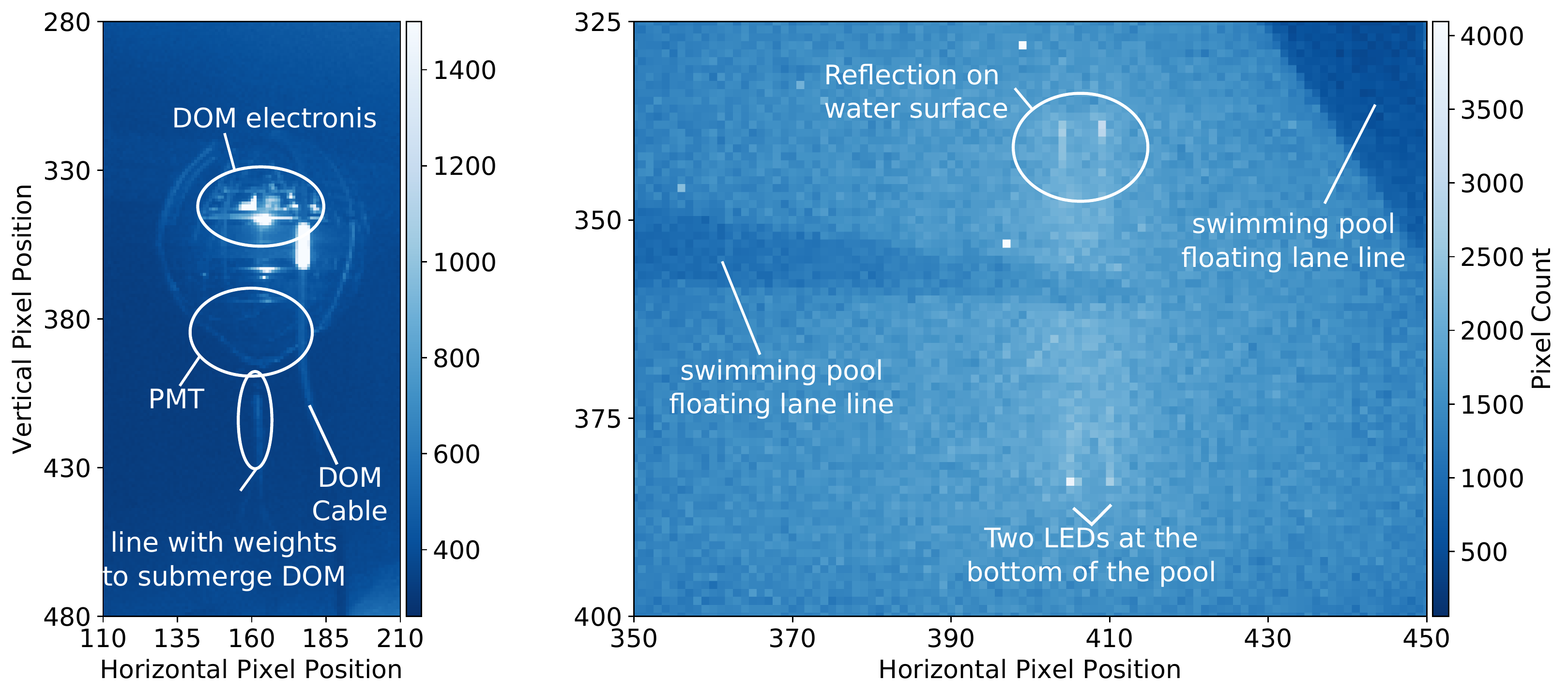}
\caption[margin=1cm]{Images captured with the upgrade camera system inside a glass pressure vessel in a swimming pool. Left: Example of a reflection photography image of an IceCube DOM at 2~m distance. Features such as the cable can be clearly seen. Right: Example of a transmission photography image of two illumination modules at a distance of 25~m. LEDs that were spaced 40~cm apart can be clearly separated on the image. The reflected image on the water surface can also be seen.}
\label{fig:Pool}
\vskip-5ex
\end{figure}

\indent To verify if the camera design and the final products adhere to the engineering requirements defined in table~\ref{tab:ERD} a series of tests is defined in table~\ref{tab:TEST}. All these tests are performed on the camera prototypes, but only a subset of these tests will be used for quality control of the serially produced cameras before integration. Additionally the alignment of each camera inside the DOMs is tested by capturing an image of a test pattern sheet at a fixed orientation and distance from within the DOM after the camera integration.\\
\indent For quality control multiple tests are conducted in the same setup for efficient testing. In the mass test setup the camera is fixed next to an illumination system at a certain distance facing a test pattern sheet inside a dark box. First communications and dark noise are tested whilst the illumination system is inactive. This trial is repeated at $-40^\circ$ C. Then the sensitivity, linearity and field of view (FOV) and lens alignment are checked with the illumination system powered on. At the bottom of the pattern sheet an LED array is used to verify the exposure time setting of the camera.\\
\begin{table}[]
    \centering
    \begin{tabular}{|l|c|}
    \hline
        \textbf{Parameter} & \textbf{Quality control and calibration tests} \\
        \hline
        \hline
        Visual inspection & Check if the camera module (incl. lens) has any visible damage.\\
        \hline
        Data Transfer & Camera settings are written to the camera registers\\
        and Communication & and read out again as multiple test images are captured.\\
        \hline
        Dark noise & Images are taken with the camera inside of a dark box\\
        & at different gain and exposure time settings.\\
        \hline
        Power Consumption & The camera and illumination system power consumption is tested \\
        & using a shunt resistor and a fast analog to digital converter setup.\\
        \hline
        Lens alignment & Images of a test pattern sheet are captured at a fixed distance. \\
        \hline
        Sensitivity & The camera response to a stable light source is measured\\
        and linearity & for different gain and exposure time settings. \\
        \hline
        Exposure time & The exposure time is measured by taking images of an array of \\
        & flashing LEDs. \\
        \hline
        LED angular & An image of the area that the illumination system  illuminates\\
        characteristics & from a given distance is used to measure the emission characteristic. \\
        \hline
        Low temperature & The dark noise, communications, exposure time, power consumption \\
        operation & and sensitivity tests are repeated at $-40 ^\circ C$. \\
        \hline
    \end{tabular}
    \caption{A table of selected tests performed on the camera system as part of quality control.}
    \label{tab:TEST}
\end{table}
\indent A full prototype operations test with a camera module and three illumination modules was performed in a 2~m deep swimming pool of the Gyeonggi Physical Education High school, a public high school for athletes in Suwon, South Korea. The camera system was installed inside of a glass pressure vessel pointing sideways at 40~cm below the water surface of the swimming pool. Small glass housings were used to contain the illumination modules. They were placed on the bottom of the pool, with the light cones pointing directly upward to minimize reflections on the water surface and floor. The critical angle for the water surface is $48.5^\circ$ compared to the $30^\circ$ opening angle of the LEDs. Several different setups were used to test the feasibility of each proposed measurement for the camera system. A selection of results is shown in Fig.~\ref{fig:Pool}. One setup was to take an image of a DOM at 2~m distance while illuminating it from the camera position. The image taken shows the clear shape of the DOM and its cable in the water. In further setups an illumination module was placed on the bottom of the pool at the horizontal distances of 10~m, 20~m, and 30~m respectively. The camera captured a significant amount of scattered light from the light cone of the illumination modules. The captured distribution and intensity of the scattered light varied with distance. In the last setup two illumination modules were placed 40~cm apart at 25~m distance from the camera. The LEDs can be clearly separated and their relative positions were measured with an uncertainty of 10~cm.

\section{Camera simulations}\label{sec:simulation}

\begin{figure}
   \centering
   \includegraphics[width=.95\linewidth]{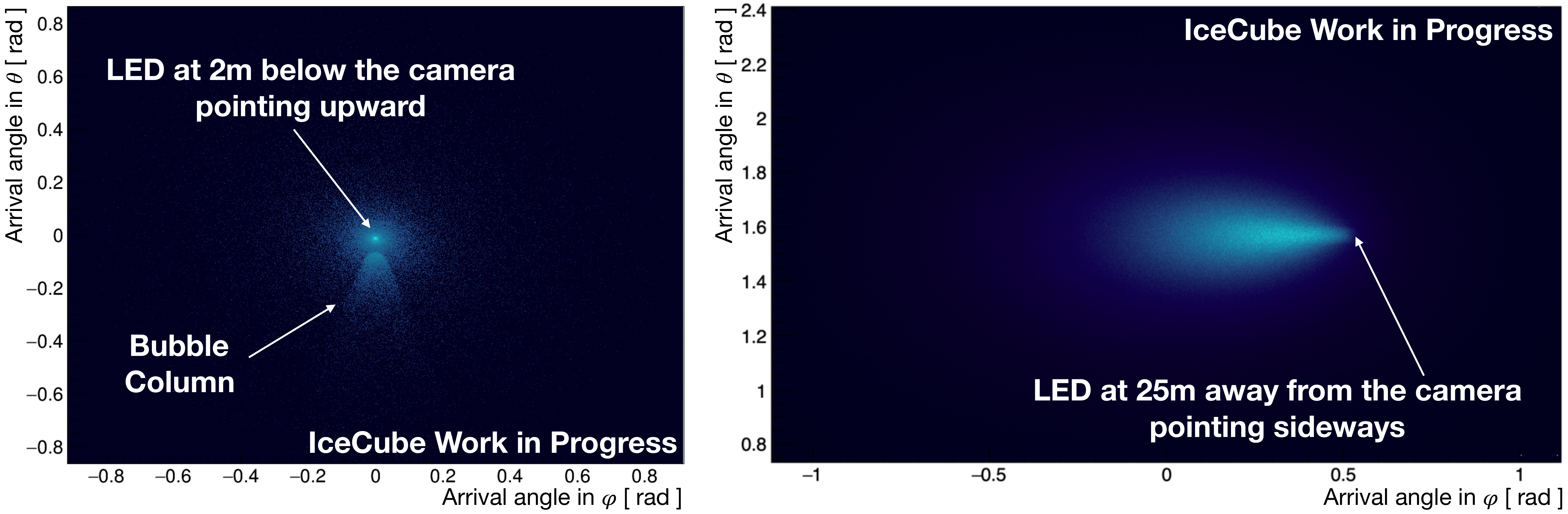}
   \caption{Expected images from the camera simulation for different configurations. Left: Refrozen hole ice with a central bubble column. Right: Scattering in the bulk ice with an LED from the adjacent string (25~m).}
   \label{fig:simsample}			
\end{figure}

\begin{figure}
   \centering
   \begin{minipage}{.98\textwidth}
    \includegraphics[width=.49\linewidth]{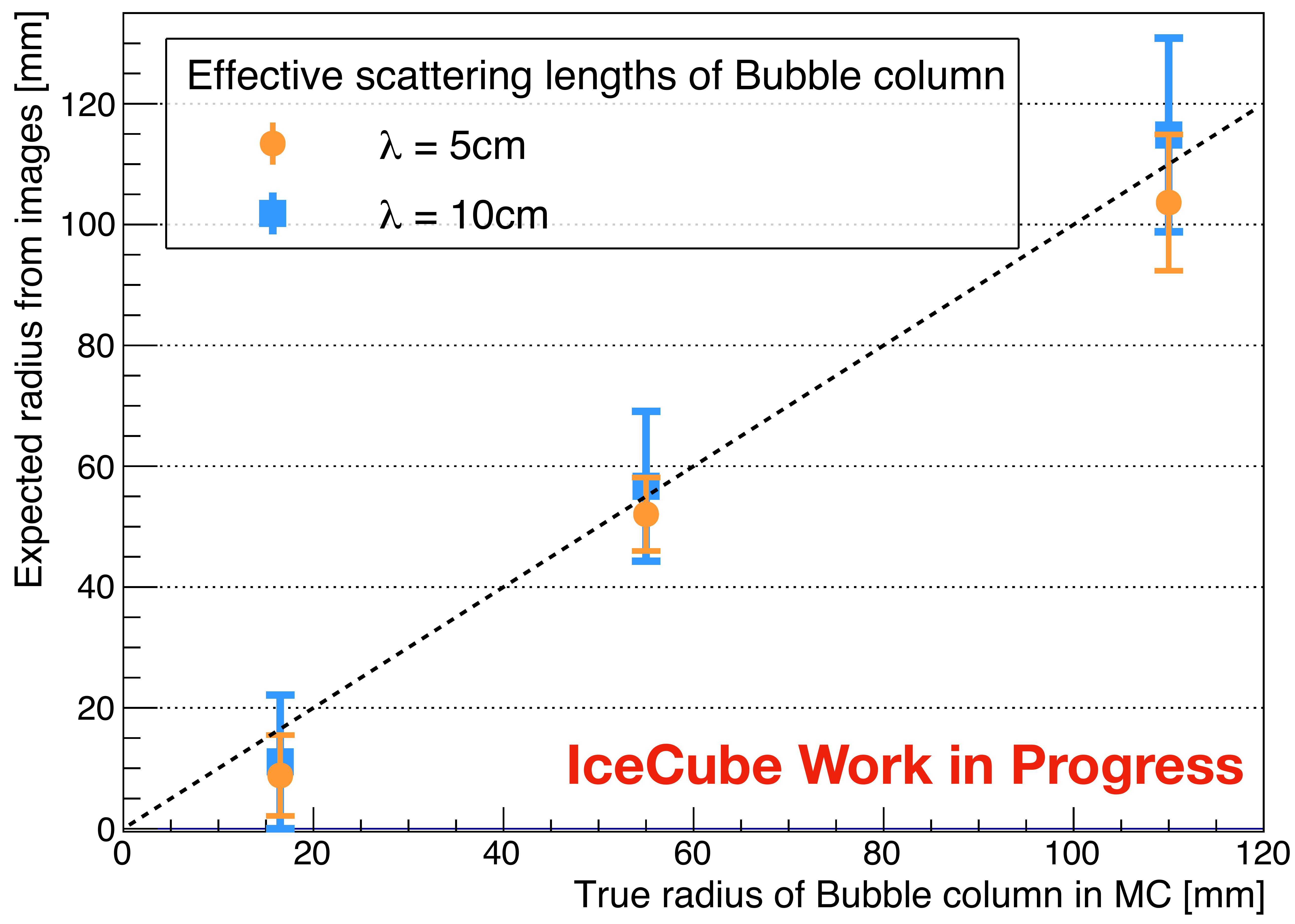}
    \includegraphics[width=.49\linewidth]{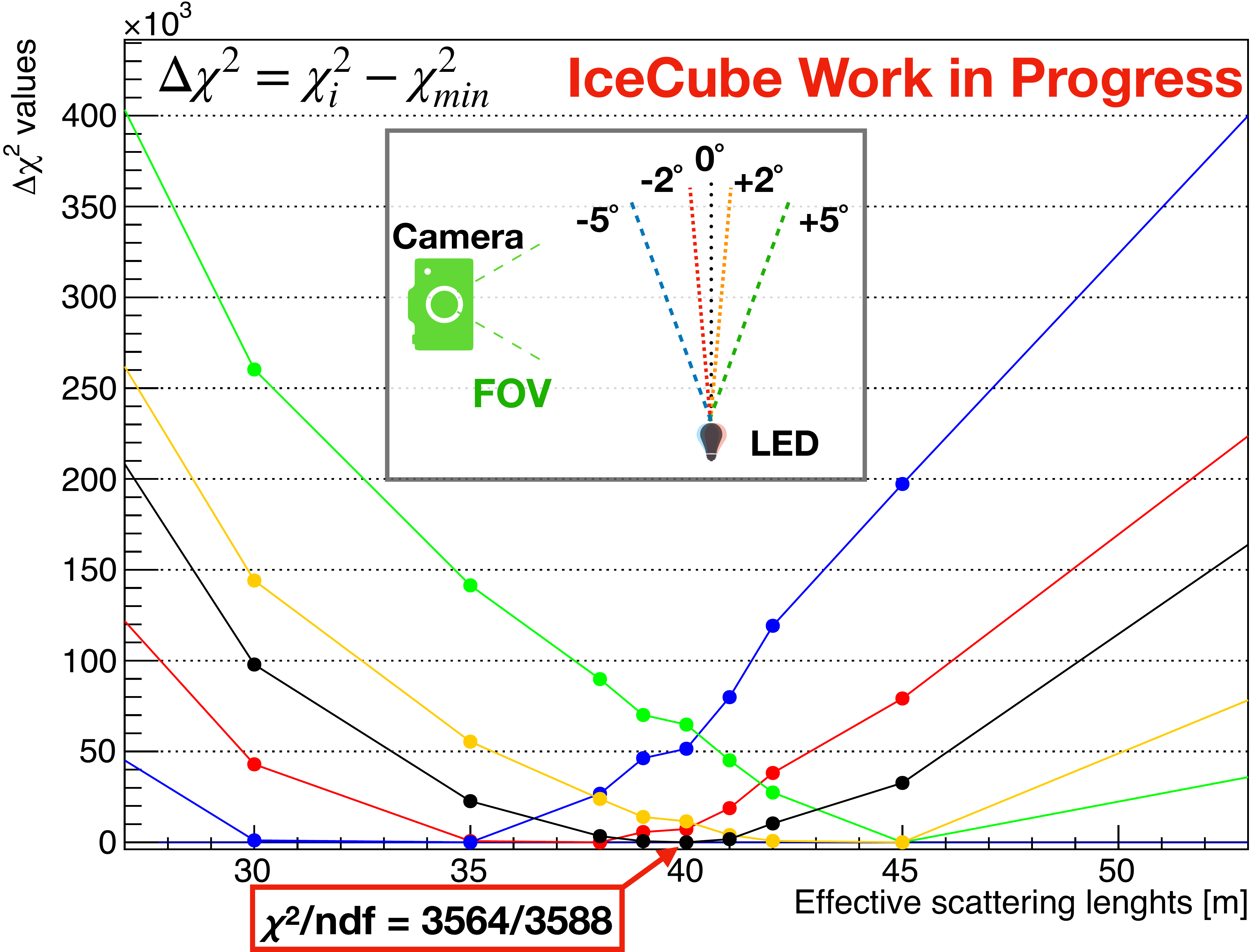}
   \end{minipage}
   \caption{Left: The bubble column radius could be measured to the $O$(1~cm) precision. The dotted line indicates the expected result from the measurements. Right: The property of the surrounding ice could be measured by comparing the images with $\Delta\chi^{2}$ method to a precision of $O$(10~m) assuming that the LED orientation is known within a few degrees. $\chi^2_{min}$ is the minimum $\chi^2$ value for a given tilt. The $\chi^{2}$ value in the red box is a typical example for the simulation with an LED tilt of 0${^{\circ}}$. Some simulated cases are outside of the axis range.}
   \label{fig:simresult}
\end{figure}

\indent To inform and optimize the design of the camera system, with regards to its ability to measure the expected light signatures, we conducted a series of simulation studies. These simulations are based on a photon propagating Monte Carlo code used in previous camera studies~\cite{ICRC2017:Gen2_camera}.\\
\indent The original simulation tool~\cite{ppc} is designed to simulate photon emissions from LEDs with a given wavelength spectrum and geometrical emission profile. It then propagates these photons within the ice medium by applying the SPICE ice model~\cite{SPICE}. The SPICE ice model and the photon propagation code is the same as is used for Cherenkov-light in standard IceCube analysis. The proposed surveys can be simulated by adjusting the positions of the camera and LED with simplified detector geometry. The results of the simulation are recorded in the form of the number of detected photons and their arrival directions at the camera. The arrival directions are used to reconstruct the images captured by the camera. The number of photons in a certain direction, representing the intensity, is converted to the expected pixel readout using a scaling factor. This factor is derived by comparing measurement in a test setup in air where a test LED shine light directly at the camera and a simulation of that exact setup. Two sample images from two different system configurations, representing the refrozen ice survey and the surrounding ice study respectively, are displayed in Fig.~\ref{fig:simsample}.\\
\indent To see whether the new camera system has the ability to survey the local ice phenomena like bubble columns, one of the simulation configurations models this situation as shown in Fig.~\ref{fig:schematic}, left. From the simulated images the column radius can be extracted by measuring size of the visible features in the image. Six different configurations were simulated in ten pseudo-experiments each, with two different scattering properties of the columns and three different column radii. The column radius was measured in each image and the results indicate a precision of a few centimetres for measuring the column radius, as shown in Fig.~\ref{fig:simresult}, left.\\
\vskip-2ex

\indent For the surrounding ice study, several different effective scattering lengths are defined as the optical properties of the ice volume between two DOMs in the same depth level of the neighboring strings in 25~m distance as illustrated in the Fig.~\ref{fig:schematic}, right. Since the orientation of LED light would be aligned randomly to the camera position, 5 different relative orientations of light were applied as a systematic variation. Images from each simulation are compared using a $\Delta\chi^{2}$ method. In the calculation of $\chi^{2}$ the images were compared to a template with high statistics in which a 40~m scattering length and no tilt was assumed. In Fig.~\ref{fig:simresult}, right the difference between the $\chi^{2}$ of each case and the minimum $\chi^{2}$ for that given tilt is shown. This plot suggest that the camera system can measure the scattering length of the surrounding bulk ice to a precision of $O$(10~m), under the assumption that the LED orientation is known within a few degrees. As several measurements will be done at each depth in the detector the effect of the uncertainty of the LED orientations will be reduced and the scattering length can then be determined with a few meter precision.\\
\vskip-3ex

\section{Conclusions}\label{sec:summary}

\indent We have developed a camera system for the IceCube Upgrade. Each optical sensor module will contain three cameras in combination with up to four illumination boards. The camera systems will be used to measure bulk ice properties and to characterize the refrozen ice in the drill hole using a series of transmission and reflection photography measurements.\\
\indent We have characterized the first prototype of the camera system and developed a series of tests for the final acceptance before integration into the optical sensor modules. A similar system~\cite{ICRC2019:SPICEcore_camera} for a scientific survey of the Antarctic ice near the IceCube detector was successfully deployed during the austral summer season of 2018/2019, demonstrating the operation principles.\\
\indent Monte Carlo simulations have been performed to determine the necessary sensitivity and characteristics of the camera system. The simulations have been calibrated to match the properties of the prototype camera system and illumination board.\\

\bibliographystyle{ICRC}

\end{document}